\documentclass[superscriptaddress, floatfix, unsortedaddress, aps, prb,
amssymb, amsmath, twocolumn]{revtex4} 

\usepackage{amsmath}
\usepackage{graphicx}

\begin{document}

\bibliographystyle{apsrev}


\newcommand{\ord}[1]{\mathcal{O}(#1)}
\newcommand{\rl}[1]{#1_{12}}
\newcommand{\eps}{\epsilon}
\newcommand{\rhat}{\hat{\mathbf{r}}_{12}}
\newcommand{\nhat}{\hat{\mathbf{n}}}
\newcommand{\rhohat}{\hat{\mathbf{\rho}}}
\newcommand{\dr}{\delta_{R}}
\newcommand{\dt}{\delta_{T}}
\newcommand{\resq}{RE${}^2_{\;\;\;}$}
\newcommand{\toluene}{C$_7$H$_8$}
\newcommand{\benzene}{C$_6$H$_6$}


\title{Coarse-grained Interaction Potentials for Anisotropic Molecules}
\author{M. Babadi}
\address{Sharif University of
Technology, Department of Physics, P.O. Box 11365-9161, Tehran,
Iran.}
\author{R. Everaers}
\address{Max-Planck-Institut f\"ur Physik komplexer
Systeme, N\"othnitzer Str. 38, 01187 Dresden, Germany}
\author{M.R. Ejtehadi}
\email{ejtehadi@sharif.edu}
\address{Sharif University of
Technology, Department of Physics, P.O. Box 11365-9161, Tehran,
Iran.}

\date{\today}

\begin{abstract}
We have proposed an efficient parameterization method for a recent
variant of the Gay-Berne potential for dissimilar and biaxial
particles and demonstrated it for a set of small organic molecules.
Compared to the previously proposed coarse-grained models, the new
potential exhibits a superior performance in close contact and large
distant interactions. The repercussions of thermal vibrations and
elasticity has been studied through a statistical method. The study
justifies that the potential of mean force is representable with the
same functional form, extending the application of this
coarse-grained description to a broader range of molecules.
Moreover, the advantage of employing coarse-grained models over
truncated atomistic summations with large distance cutoffs has been
briefly studied.
\end{abstract}

\maketitle

\section{Introduction}
The development of accurate, reliable and computationally efficient
interaction models is the main activity of molecular modeling. The
need to attain larger simulated time scales and the excessive
complexity of a wide range of molecular systems (e.g. biomolecular)
has emphasized the factor of computation efficiency as a dominant
deliberation in choosing the appropriate interaction model for
molecular simulations. In particular, grouping certain atoms into
less detailed interaction sites, known as "coarse-graining", in one
way of achieving such efficiency.

Various coarse-grained (CG) approaches have been recently developed
with such goal in mind~\cite{GB, BFZ98, Shelley, Izvekov}. The
implementation of coarse-graining models is usually divided into two
distinct stages. The first is a partitioning of the system into the
larger structural units while the second stage is the construction
of an effective force field to describe the interactions between the
CG units. Typically, CG potentials of a pre-defined analytical form
are parameterized to produce average structural properties seen in
atomistic simulations. Such analytical forms are chosen in a way to
describe the governing interaction between the CG
units~\cite{Shelley}. The parameterizations are usually based on
matching samples of potentials of mean force~\cite{Shelley, Meyer},
inverse Monte Carlo data~\cite{Murtola} or certain atomistic
potentials characteristics~\cite{BFZ98}. The main concern of the
present work is parameterizing a CG force field for the short-range
attractive and repulsive interactions between ellipsoidal molecules
and groups, based on atomistic potential sampling and potential of
mean force.

In molecular simulations, short-range attractive and repulsive
interactions are typically represented using Lennard-Jones(6-12)
potentials~\cite{AllenTildesley, FrenkelSmit}:
\begin{equation}\label{eq:ULJ}
U_{LJ}(r; i,j) = 4\epsilon_{ij} \left[ \left(\frac{\sigma_{ij}}
r\right)^{12} -
       \left(\frac{\sigma_{ij}} r\right)^{6}
\strut\right]
\end{equation}
where $\sigma_{ij}$ and $\epsilon_{ij}$ are the effective
heterogeneous interaction radius and well-depth between particles of
type $i$ and~$j$ respectively and $r$ is the inter-particle
displacement. While the $r^{-6}$ part has a physical origin in
dispersion or van der Waals interactions, the $r^{-12}$ repulsion is
chosen for mathematical convenience and is sometimes replaced by
exponential terms as well. For large molecules, the exact evaluation of the
interaction potential of this type involves a computationally
expensive double summation over the respective (atomic) interaction
sites:
\begin{equation}\label{eq:UMacroSum}
U_{int}(\mathcal{M}_1, \mathcal{M}_2) =
\sum_{i\in\mathcal{M}_1}\sum_{j\in\mathcal{M}_2} U_a(r_{ij}; i, j)
\end{equation}
where $\mathcal{M}_1$ and $\mathcal{M}_2$ denote the interacting
molecules and $U_a(\cdot)$ is the atomic interaction potential, e.g.
Eq.~(\ref{eq:ULJ}). In practice, a large distant interaction cutoff
accompanied by a proper tapering is used to reduce the computation
cost. More sophisticated and efficient summation methods such as
Ewald summation and the Method of Lights are also widely
used~\cite{Leach}.

As an alternative approach, Gay and Berne~\cite{GB} proposed a more
complicated single-site CG interaction potential (in contrast to
sophisticated summation techniques) for uniaxial rigid molecules
which was generalized to dissimilar and biaxial particles later by
Berardi {\it et al} as well~\cite{BFZ98}. We will refer to this
potential as the biaxial-GB in the rest of this article.

In response to the criticism of the unclear microscopic
interpretation of the GB potential~\cite{Perram96}, we have recently
used results from colloid science~\cite{Hunter} to derive an
approximate interaction potential based on the Hamaker
theory~\cite{Hamaker} for mixtures of ellipsoids of arbitrary size
and shape, namely the \resq potential~\cite{EE03}. Having a
parameter space identical to that of Berardi, Fava and
Zannoni~\cite{BFZ98}, the \resq potential agrees significantly
better with the numerically evaluated continuum approximation of
Eq.~(\ref{eq:UMacroSum}), has no unphysical large distant limit and
avoids the introduction of empirical adjustable parameters.

In an anisotropic coarse-grained model, a molecule $\mathcal{M}$ is
treated like a rigid body. Neglecting the atomic details, each
molecule is characterized by a center separation $\mathbf{r}$ and a
transformation operator (a unitary matrix $\mathbf{A}$ or a unit
quaternion $\mathbf{q}$) describing its orientation.

In the first section of the article, we briefly introduce the \resq
potential followed by a review of the biaxial-GB potential and
Buckingham(exp-6) atomistic model. The Buckingham(exp-6) potential
is used in the MM3 force field~\cite{MM3} and will serve as the
atomistic model potential for parameterizations. The second section
describes a parameterization method which has been demonstrated for
a few selected molecules, followed by an exemplar comparison between
the \resq and the biaxial-GB potential. We will study the
repercussion of internal vibrations, in contrast to the usually
assumed proposition of the ideal stiffness~\cite{GB, BFZ98}, and
propose an error analysis method to define trust temperature regions
for single site potentials. Finally, we will show that the potential
of mean force (PMF) is representable with the same functional form
for a wide range of temperatures.

\section{Atomistic and Single-site Non-bonded Potentials}
In the MM3 force field~\cite{MM3}, the van der Waals interaction is
described in terms of Buckingham(exp-6) potential which is an exponential
repulsive accompanied by a $r^{-6}$ attractive term:
\begin{equation}\label{eq:mm3a}
U^{MM3}(r_{ij}; i, j) = \epsilon_{ij}\left(A
e^{-B\sigma_{ij}/r_{ij}}-C\left(\frac{\sigma_{ij}}{r_{ij}}\right)^6\right)
\end{equation}
where A, B and C are fixed empirical constants while $\sigma_{ij}$
and $\epsilon_{ij}$ are heterogeneous interaction parameters specific
to the interacting particles. Usually, Lorenz and Berthelot
averaging rules are used to define heterogeneous interaction
parameters in terms of the homogeneous ones, i.e. $\sigma_{ij} =
(\sigma_{i} + \sigma_{j})/2$ and $\epsilon_{ij} =
\sqrt{\epsilon_{i}\epsilon_{j}}$. The hard core repulsion is usually
described via a $r^{-12}$ term with an appropriate energy switching:
\begin{equation}\label{eq:mm3r}
U_{HC}^{MM3}(r_{ij}; i, j) = \gamma
\left(\frac{\sigma_{ij}}{r_{ij}}\right)^{12}
\end{equation}
where $\gamma$ is defined in a way to provide continuity at the
switching distance. The interaction energy between two arbitrary molecules
is trivially the pairwise double summation over all of the interaction
sites, i.e. Eq.~(\ref{eq:UMacroSum}).

The dissimilar and biaxial Gay-Berne potential (biaxial-GB) is a
widely used single-site model proposed by Berardi et
al.~\cite{BFZ98} which is an extension of the original uniaxial
description~\cite{GB} to biaxial molecules and heterogeneous
interactions. Based on the original Gay and Berne concept, the
biaxial-GB is a shifted Lennard-Jones(6-12) interaction between two
biaxial Gaussian distribution of interacting sites. In this
coarse-grained model, each molecule is described by two diagonal
characteristic tensors (in the principal basis of the molecule)
$\mathbf{S}$ and~$\mathbf{E}$, representing the half radii of the
molecule and the strength of the pole contact interactions,
respectively. As mentioned earlier, the orientation of a molecule is
described by a center separation vector $\mathbf{r}$ and a unitary
operator~$\mathbf{A}$, revolving
the lab frame to the principal frame of the molecule.\\

The biaxial-GB description for the interaction between two molecules
with a center separation of $\rl{\mathbf{r}}= \mathbf{r}_2 -
\mathbf{r}_1$ and respective orientation tensors $\mathbf{A}_1$ and
$\mathbf{A}_2$ is defined as:
\begin{multline}\label{eq:GBdef}
U^{GB}_{A,R}(\rl{\mathbf{r}},\mathbf{A}_1,\mathbf{A}_2)=\\
4\epsilon_0\rl{\eta}^{\nu}\rl{\chi}^{\mu}
\Bigg[\bigg(\frac{\sigma_c}{\rl{h}+\sigma_c}\bigg)^{12}-
\bigg(\frac{\sigma_c}{\rl{h} + \sigma_c}\bigg)^{6}\Bigg]
\end{multline}\\
where $\epsilon_0$ and $\sigma_c$ are the energy and length scales,
$\rl{\eta}$ and $\rl{\chi}$ are purely orientation dependant terms~\cite{BFZ98}
and $\rl{h}$ is the the least contact distance between the two
ellipsoids which are defined by the diagonal covariance tensor of the assumed
Gaussian distributions. The orientation dependant terms ($\rl{\eta}$
and $\rl{\chi}$) describe the anisotropy of the molecules.\\

We have recently proposed a single-site potential, namely
\resq~\cite{EE03} giving the approximate interaction energy between
two hard ellipsoids in contrast to the tradition of the Gaussian
clouds, initiated by Gay and Berne~\cite{GB}. The orientation
dependence of the \resq potential fall at large distances, reducing
asymptotically to the interaction energy of two spheres. Moreover,
it gives a more realistic intermediate and close contact interaction
using a heuristic interpolation of the Deryaguin
expansion~\cite{Deryaguin, EE03}. Being a shifted
Lennard-Jones(6-12) potential, the biaxial-GB fails to exhibit the
correct functional behavior for large molecules~\cite{EE03}. The
attractive and repulsive contributions of the \resq potential are
respectively:
\begin{subequations}\label{eq:re2}
\begin{multline}\label{eq:re2A}
U_A^{RE^2}(\mathbf{A}_1, \mathbf{A}_2, \rl{\mathbf{r}}) =-\frac{A_{12}}{36}\Big(1+
3\rl{\eta}\rl{\chi}\frac{\sigma_c}{\rl{h}}\Big)\times\\
\prod_{i=1}^2\prod_{e=x,y,z}
\Bigg(\frac{\sigma^{(i)}_e}{\sigma^{(i)}_e+h_{12}/2}\Bigg)
\end{multline}
\begin{multline}\label{eq:re2R}
U_R^{RE^2}(\mathbf{A}_1, \mathbf{A}_2, \rl{\mathbf{r}})=\frac{A_{12}}{2025}
\Big(\frac{\sigma_c}{\rl{h}}\Big)^6\Big(1+
\frac{45}{56}\rl{\eta}\rl{\chi}\frac{\sigma_c}{\rl{h}}\Big)\times\\
\prod_{i=1}^2 \prod_{e=x,y,z}
\Bigg(\frac{\sigma^{(i)}_e}{\sigma^{(i)}_e+h_{12}/60^{\frac{1}{3}}}\Bigg)
\end{multline}
\end{subequations}\\
where $A_{12}$ is the Hamaker constant (the energy scale),
$\sigma_c$ is the atomic interaction radius and $\sigma_x^{(i)}$,
$\sigma_y^{(i)}$ and $\sigma_z^{(i)}$ are the half-radii of $i$th
ellipsoid (i=1,2). The terms $\rl{\eta}$, $\rl{\chi}$ and $\rl{h}$ are
defined in parallel to the biaxial-GB model and thus, are described
in terms of the same characteristic tensors.\\

The structure tensor $\mathbf{S}_i$ and the relative potential well depth
tensor $\mathbf{E}_i$ are diagonal in the principal basis of $i$th molecule
and are defined as:
\begin{subequations}
\begin{equation}
\mathbf{S}_i = \textrm{diag}\{\sigma_x^{(i)}, \sigma_y^{(i)},
\sigma_z^{(i)}\}
\end{equation}
\begin{equation}
\mathbf{E}_i = \textrm{diag}\left\{E_x^{(i)}, E_y^{(i)},
E_z^{(i)}\right\}
\end{equation}
\end{subequations}
where $E_x^{(i)}$, $E_y^{(i)}$ and $E_z^{(i)}$ are dimensionless energy scales
inversely proportional to the potential well depths of the respective orthogonal
configurations of the interacting molecules ({\bf aa}, {\bf bb} and {\bf cc},
Table~\ref{tab:ortho}). For large molecules with uniform constructions, it has
been shown~\cite{EE03} that the energy parameteres are approximately representable
in terms of the local contact curvatures using the Deryaguin expansion~\cite{Deryaguin}:
\begin{equation}\label{eq:derya}
\mathbf{E}_i = \sigma_c \textrm{diag}\left\{\frac{\sigma_x}{\sigma_y
\sigma_z}, \frac{\sigma_y}{\sigma_x \sigma_z},
\frac{\sigma_z}{\sigma_x \sigma_y}\right\}
\end{equation}
The assumptions leading to these estimations are not valid for the
studied small organic molecules. Therefore, we will cease to impose
further suppositions and take these three scales as independent characteristics
of a biaxial molecule. Computable expressions for the orientation dependent factors
of the \resq potential ($\rl{\eta}$ and $\rl{\chi}$) among with the Gay-Berne
approximation for $\rl{h}$ has been given in the Appendix~(\ref{sec:orientation}).

\section{Parameterization for Arbitrary Molecules}
\subsection{The Principal Basis and The Effective Center of Interaction}
Associating a biaxial ellipsoid to an arbitrary molecule, one must
define an appropriate principal basis and a center of interaction for it
beforehand, according to the used coarse-grained model. Although there's
no trivial solution to this problem, the centroid and the eigenbasis of the
geometrical inertia tensor of the molecule are promising candidates and may
be taken as suitable initial guesses as they yield to the correct solution at
least for the molecules with perfect symmetry. For a molecule consisting of
$N$ particles, the centroid is defined as:
\begin{equation}
\mathbf{r}_c = \frac{\sum_{i=1}^{N}\mathbf{r}_i}{N}
\end{equation}
and the principal basis is the eigenbasis of the geometrical inertia
tensor $\mathbf{I}_g$ given by:
\begin{equation}
\mathbf{I}_g =
\sum_{i=1}^{N}(r_i^2\mathbf{1}-\mathbf{r_i}\otimes\mathbf{r}_i)
\end{equation}
where $\mathbf{r}_i$ is the position of $i$th atom.\\

The most general parameter space of the \resq potential contains the energy and
length scales, the characteristic tensors and the parameters specifying the relative
orientation of the ellipsoids to the molecules. In order to overcome the degeneracy
of the parameter space and to guarantee the rapid convergence of the optimization
routines, a two-stage parameterization is proposed. In the first stage, the center and
principal frame of the molecule will be fixed at the centroid and the eigenbasis of the
inertia tensor. A preliminary optimization in the reduced parameter space yields
to an approximate parameterization. In the second stage, the results of the first stage
will be taken as the initial guess, followed by an optimization in the unconstrained
variable space. This two-stage parameterization will theoretically result in
superior results for molecules with imperfect symmetries.

\subsection{Sampling and Optimization}
Physical and symmetrical considerations lead to the proposition that
a sampling of the pole contact interactions between two biaxial particles is
essentially sufficient to reproduce the interaction for all configurations.
There are 18 different orthogonal approaching configurations (pole contacts)
between two dissimilar and biaxial particles (Table~\ref{tab:ortho}).
Based on physical grounds, we optimize the parameter space for the important
characteristics of the sampled orthogonal energy profiles, i.e. potential well
depth, potential well distance, the width of well at half depth and the soft contact
distance. This parameterization fashion is guaranteed to produce a satisfactory
reconstruction of the most crucial region of interaction.\\

The geometry of the molecules were initially optimized using TINKER
molecular modeling package~\cite{TINKER} with the MM3 force field. We
have used the same force field to sample the interaction energy for
the orthogonal configurations. \\

Given a parameter tuple $\mathbf{p}$, we denote the potential well depth,
well distance, well width at half depth and the soft contact distance for $i$th
orthogonal configuration predicted by the \resq potential by $U_m(i;\mathbf{p})$,
$R_m(i;\mathbf{p})$, $W(i;\mathbf{p})$ and $R_{sc}(i;\mathbf{p})$ respectively.
The same potential well specifications calculated from the atomistic sum is
denoted by scripted letters. An appropriate cost function is:
\begin{multline}
\Omega(\mathbf{p})=\frac{1}{\Omega_0}\sum_{i=1}^{N_{\dagger}}
e^{-\beta\mathcal{U}_m(i)}\Bigg[
w_{U_m}\bigg(\frac{U_m(i;\mathbf{p})-\mathcal{U}_m(i)}{\mathcal{U}_0}\bigg)^2+\\
w_{R_m}\bigg(\frac{R_m(i;\mathbf{p})-\mathcal{R}_m(i)}{\mathcal{R}_0}\bigg)^2
+w_W\bigg(\frac{W(i;\mathbf{p})-\mathcal{W}(i)}{\mathcal{W}_0}\bigg)^2+\\
w_{R_{sc}}\bigg(\frac{R_{sc}(i;\mathbf{p})-\mathcal{R}_{sc}(i)}{\mathcal{R}_{sc0}}\bigg)^2\Bigg]
\end{multline}
where $\Omega_0$ is a normalization factor:
\begin{equation}
\Omega_0 = 4(w_{U_m}^2 + w_{R_m}^2 + w_{W}^2 +
w_{R_{sc}}^2)^{\frac{1}{2}}\sum_{i=1}^{N_{\dagger}}e^{-\beta
\mathcal{U}_m(i)}.
\end{equation}
We have chosen $\mathcal{U}_0$ as
$\min\left\{|\mathcal{U}_m(i)|\right\}$ and $\mathcal{R}_0$,
$\mathcal{W}_0$ and $\mathcal{R}_{sc0}$ as
$\min\left\{\mathcal{W}_m(i)\right\}$ based on physical
considerations. $N_{\dagger}$ is the number of orthogonal profiles
(12 and 18 for homogeneous and heterogeneous interactions
respectively) and $(w_U, w_R, w_W, w_{R_{sc}})$ are fixed error
partitioning factors for different terms, set to $(1.0, 3.0, 2.5,
1.0)$ in order to emphasize on the structural details. We have also
included a fixed error weighting according to the Boltzmann
probability of the appearance of the corresponding profiles. One
expects higher amplitude of relative appearance for orientations
with deeper wells, which justifies the requisite of higher
contribution in the cost function. We have also chosen $\beta$ as
$1/\left\langle|\mathcal{U}_m(i)|\right\rangle$ in
order to avoid deep submergence of the lower energy orientations.\\

Further implications such as matching the large distance behavior will be
regarded as constraints on the parameter space, leaving the defined cost
function unchanged.\\

The nonlinear optimization procedure consists of a preliminary Nelder-Mead Simplex
search followed by a quasi-Newton search with BFGS Hessian updates~\cite{NumOpt}.
The whole parameterization routine is coded in MATLAB/Octave and is freely
available~\cite{RE2CLib}. The procedures of sampling and parameterization are purely
automated and requires only a Cartesian input file. The interaction parameters for the
homogeneous interaction of a set of small prolate and oblate organic molecules has been
provided in Table~(\ref{tab:param}). It is noticed that the provided half radii agree
significantly better with the molecular dimensions compared to the biaxial-GB
parameterizations~\cite{BFZ98}, reflecting the precise microscopic interpretation of
the \resq potential.

\subsection{Large Distance Analysis}
\begin{figure*}
\center
\includegraphics[bb=40 195 560 600, scale=0.4]{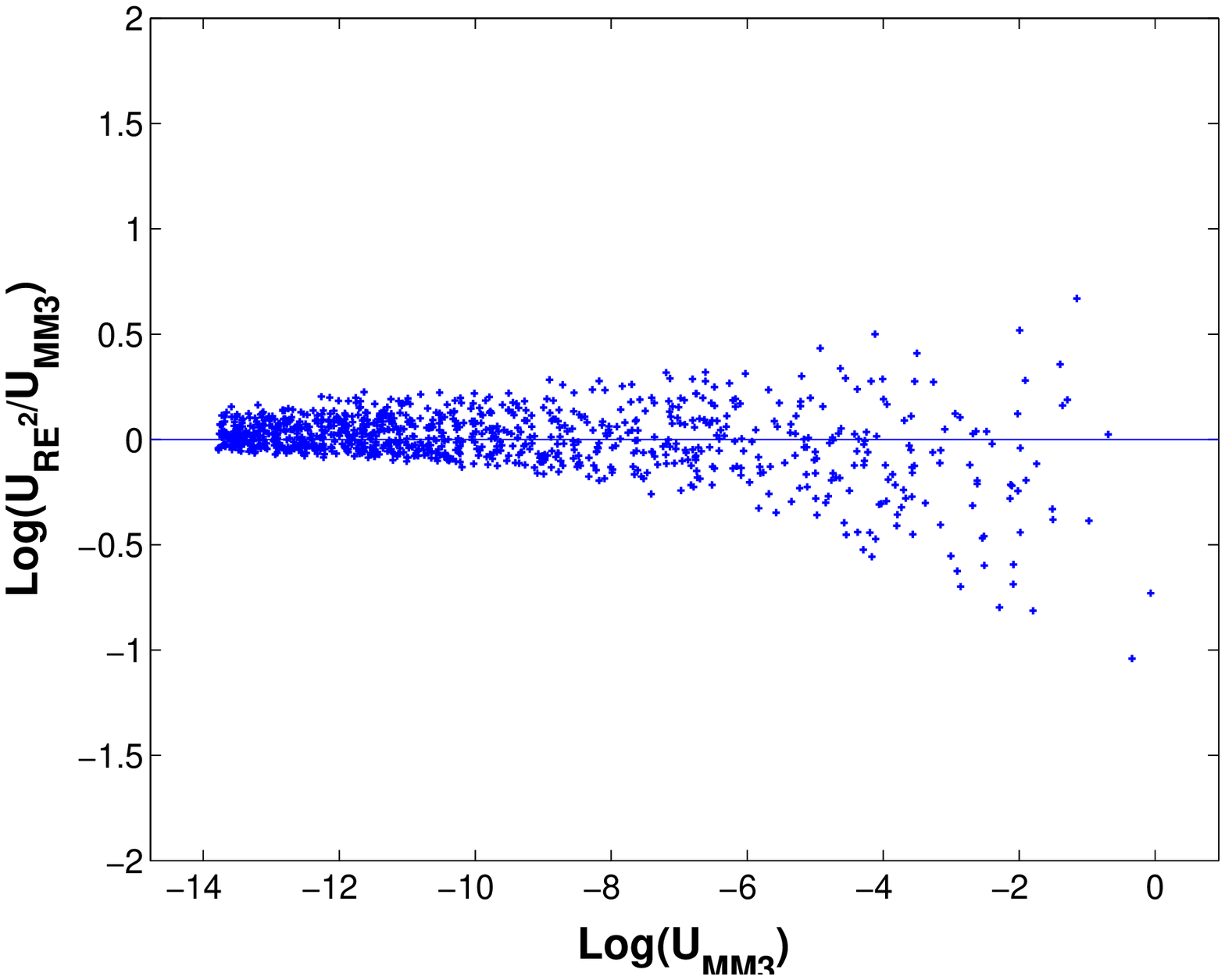}
\includegraphics[bb=40 195 560 600, scale=0.4]{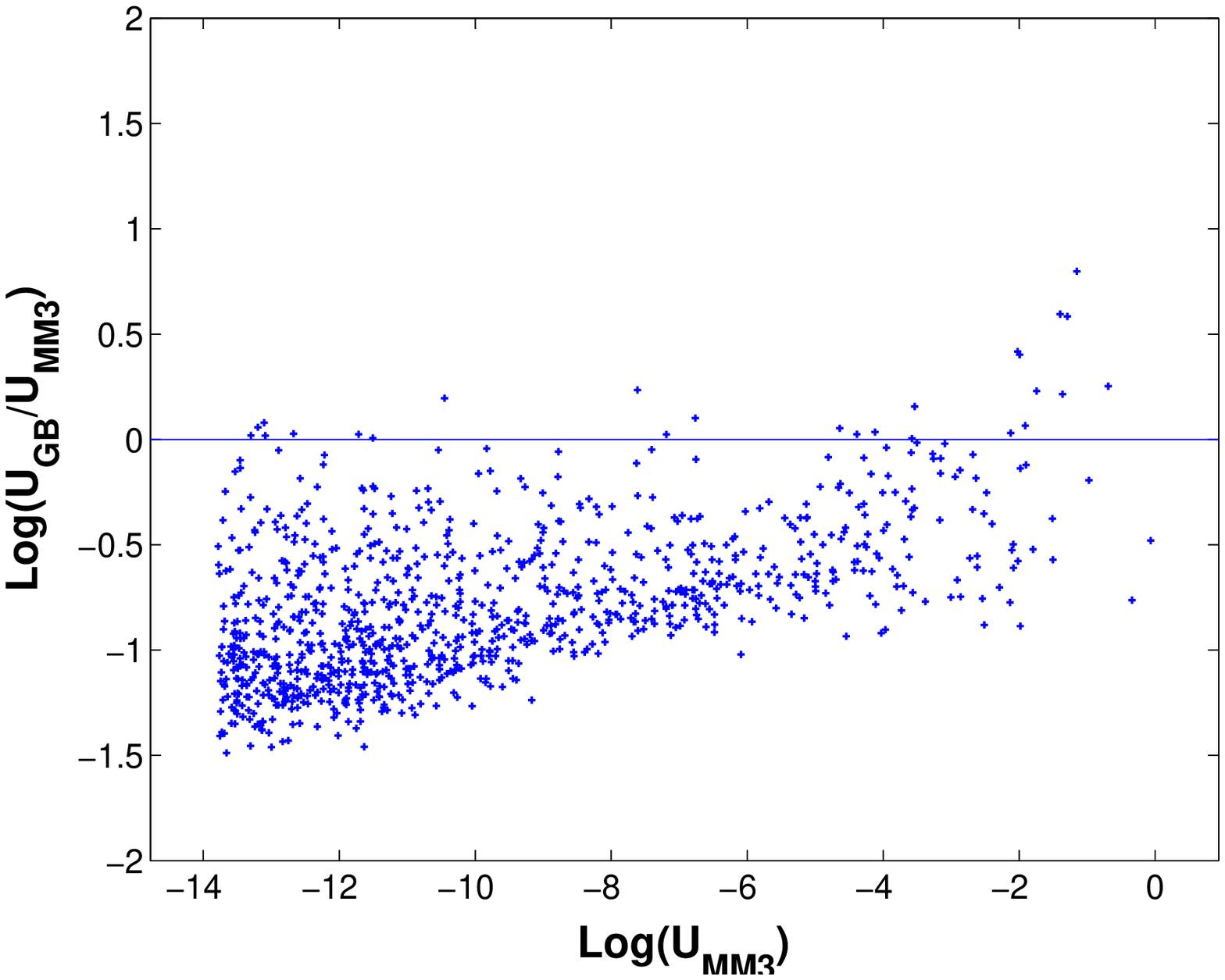}
\caption{A comparison between the \resq and biaxial-GB potentials
for the homogeneous interaction of the pair perylene in a set of
uniform random center separations in the range [5\AA, 50\AA] along
with random rotations. The Gay-Berne approximation has been used for
the least contact distance. (a) A log-log plot of $U_{RE^2}/U_{MM3}$
against $U_{MM3}$ (Mean=-0.002, SD=0.08) (b) A log-log plot of
$U_{GB}/U_{MM3}$ against $U_{MM3}$ (Mean=-0.87, SD=0.54)}
\label{fig:loglog}
\end{figure*}

The cost function defined in the previous section focuses on close contact regions only.
In order to achieve the correct large distant limit as well, we will constrain the
variable space by matching the asymptotic behavior
of the \resq potential with the atomistic summation. The asymptotic behavior of the
\resq potential is described as:
\begin{equation}\label{eq:re2limita}
\lim_{\rl{r}\rightarrow\infty}\rl{r}^6U_{RE^2}(\rl{\mathbf{r}},\mathbf{A}_1,
\mathbf{A}_2)=-\frac{16}{9}\rl{A}\det[\mathbf{S}_1]\det[\mathbf{S}_2]
\end{equation}\\
The atomistic summation defined by Eq.~(\ref{eq:UMacroSum}) and
Eq.~(\ref{eq:mm3a}) exhibits the same asymptotic behavior, which
together with Eq.~(\ref{eq:re2limita}) results in the relation:
\begin{equation}\label{eq:re2limit}
\rl{A}\det[\mathbf{S}_1]\det[\mathbf{S}_2]=\frac{9}{16}\sum_{i\in
\mathrm{A}}\sum_{j\in \mathrm{B}}\epsilon_{ij}\sigma_{ij}^6
\end{equation}\\
The summation appearing in right hand side is most easily evaluated
by a direct force field parameter lookup. Applying such a constraint
guarantees the expected large distant behavior while leads to a
faster parameterization, reducing the dimensions of the variable
space. Unconstrained optimization routines are still applicable as
one may solve Eq.~(\ref{eq:re2limit}) for $A_{12}$ explicitly. A graphical
comparison between the biaxial-GB and the \resq potential has been given for
the homogeneous interaction of the pair Perylene~\cite{Mols} has been sketched
in Fig.~\ref{fig:loglog}. The large distance convergence of the \resq potential
is noticed in contrast to the divergent behavior of the biaxial-GB potential,
which is due to the non-vanishing orientation dependent pre-factors.
Although the energy contribution is small at this limit, it is not generally
negligible, e.g. the large distant separability of the orientation dependence of
the model potential alters the nature of the phase diagram and the long range
order of a hard rod fluid in general~\cite{Cuesta}.

\subsection{Heterogeneous Interactions}
\begin{figure*}
\center
\includegraphics[bb=40 195 560 600, scale=0.35]{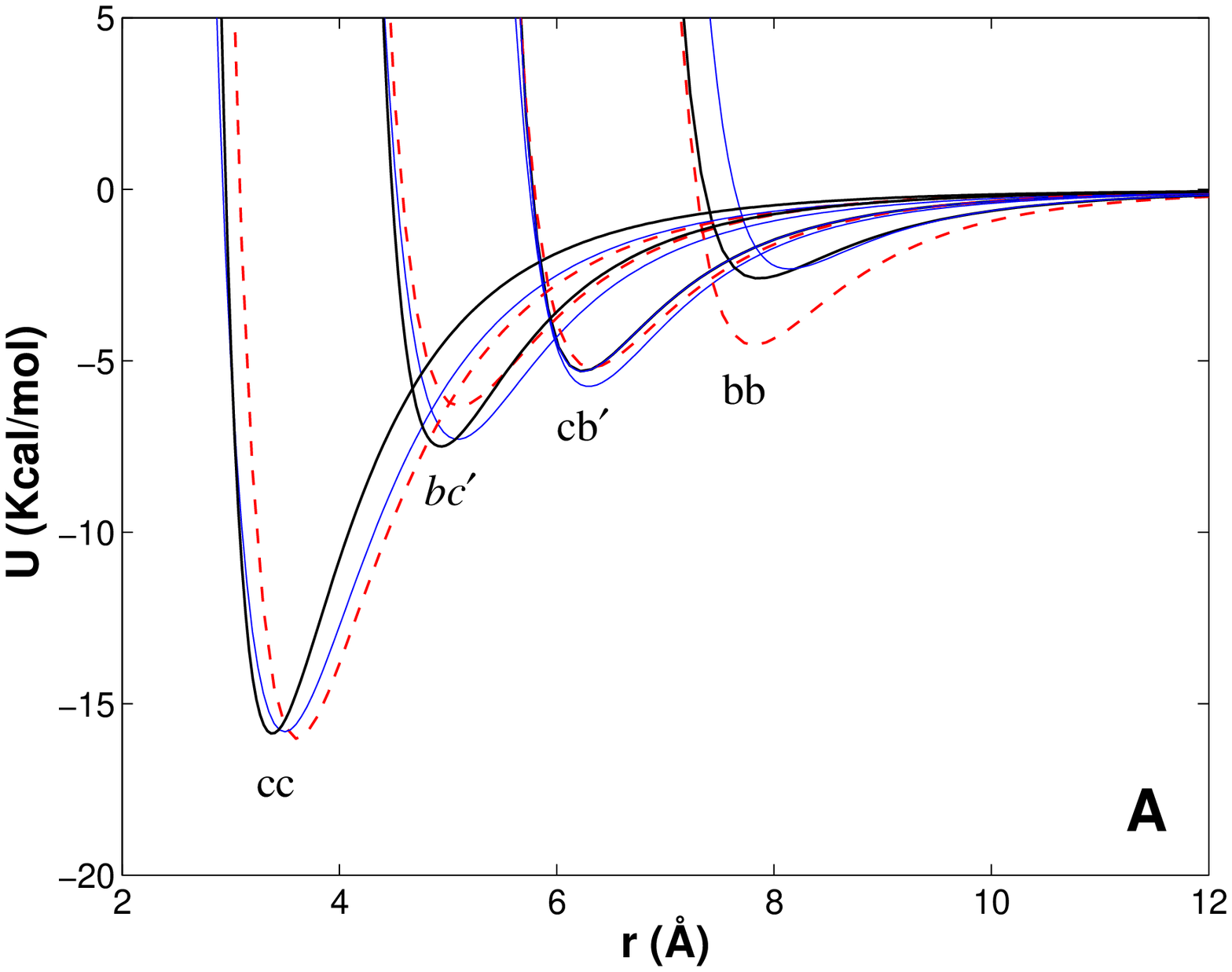}
\includegraphics[bb=40 195 560 600, scale=0.35]{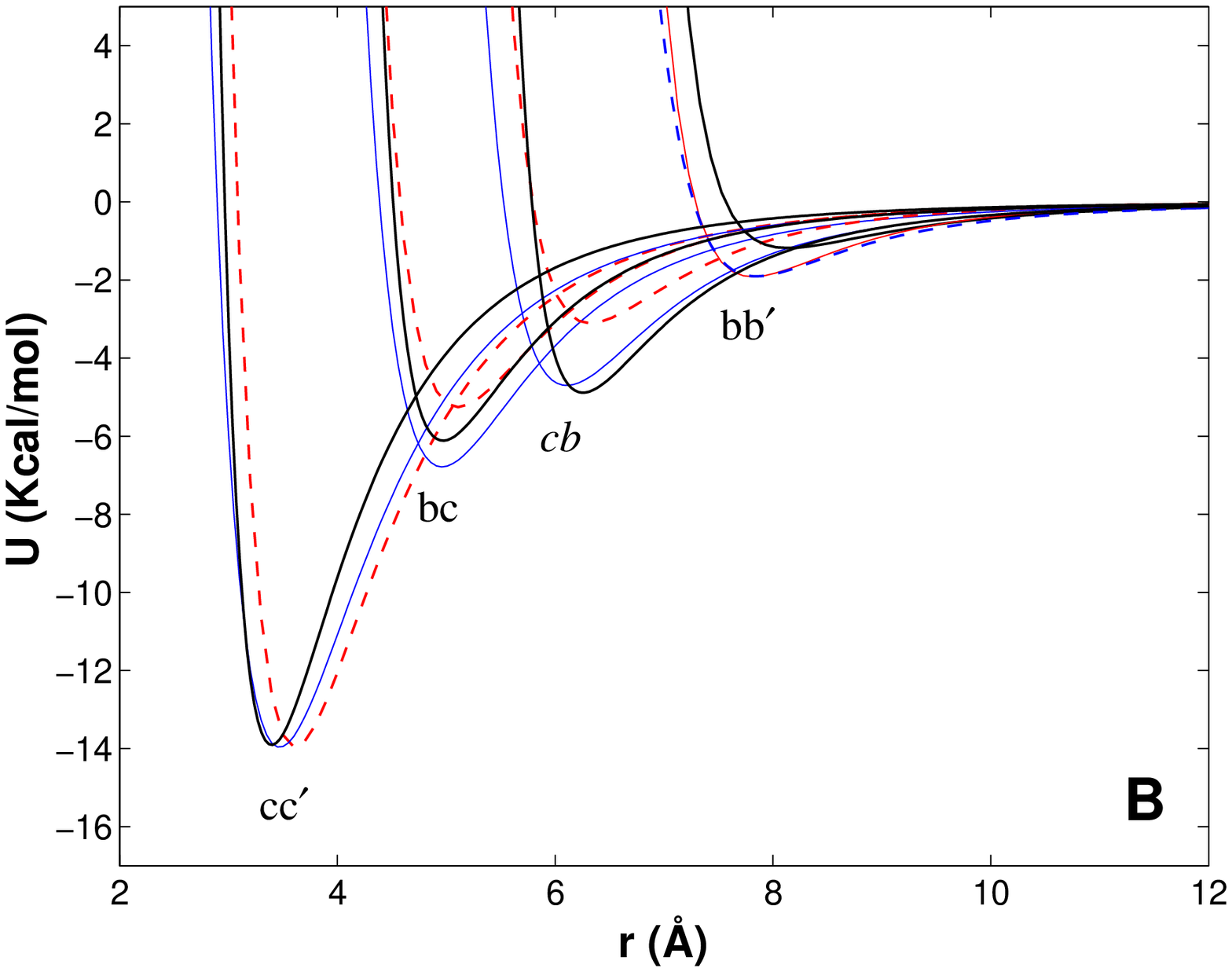}
\includegraphics[bb=40 195 560 600, scale=0.35]{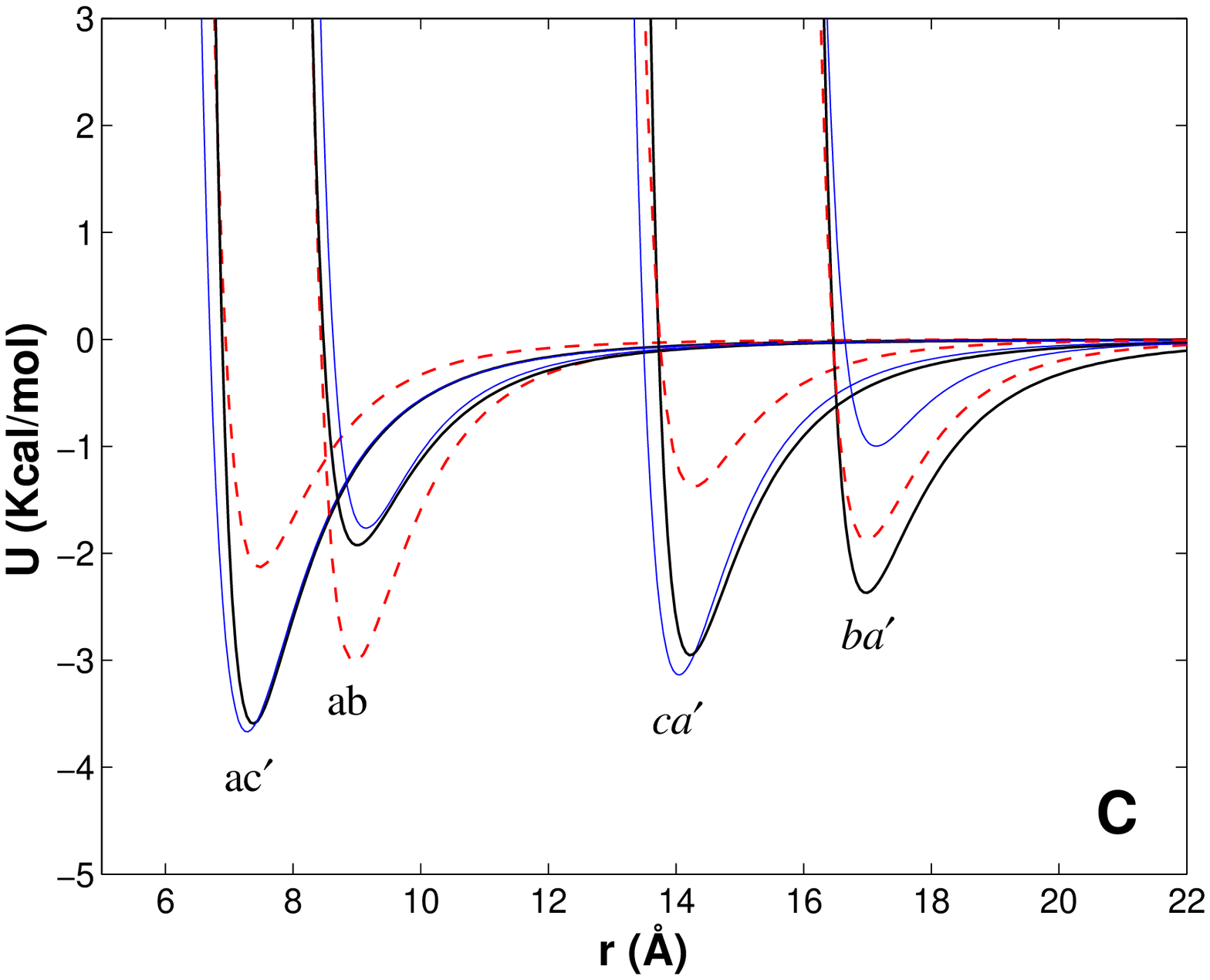}
\includegraphics[bb=40 195 560 600, scale=0.35]{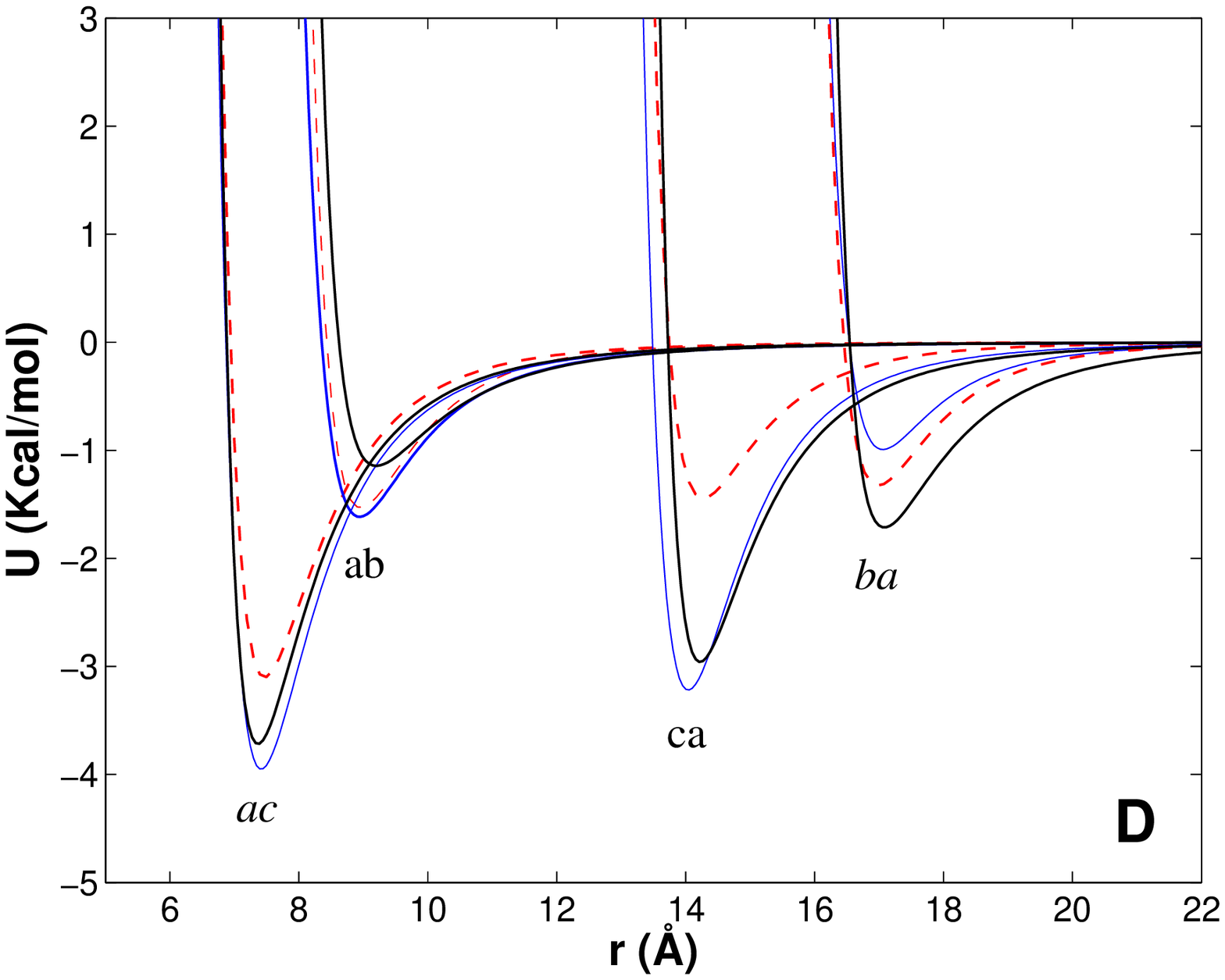}
\includegraphics[bb=40 195 560 600, scale=0.35]{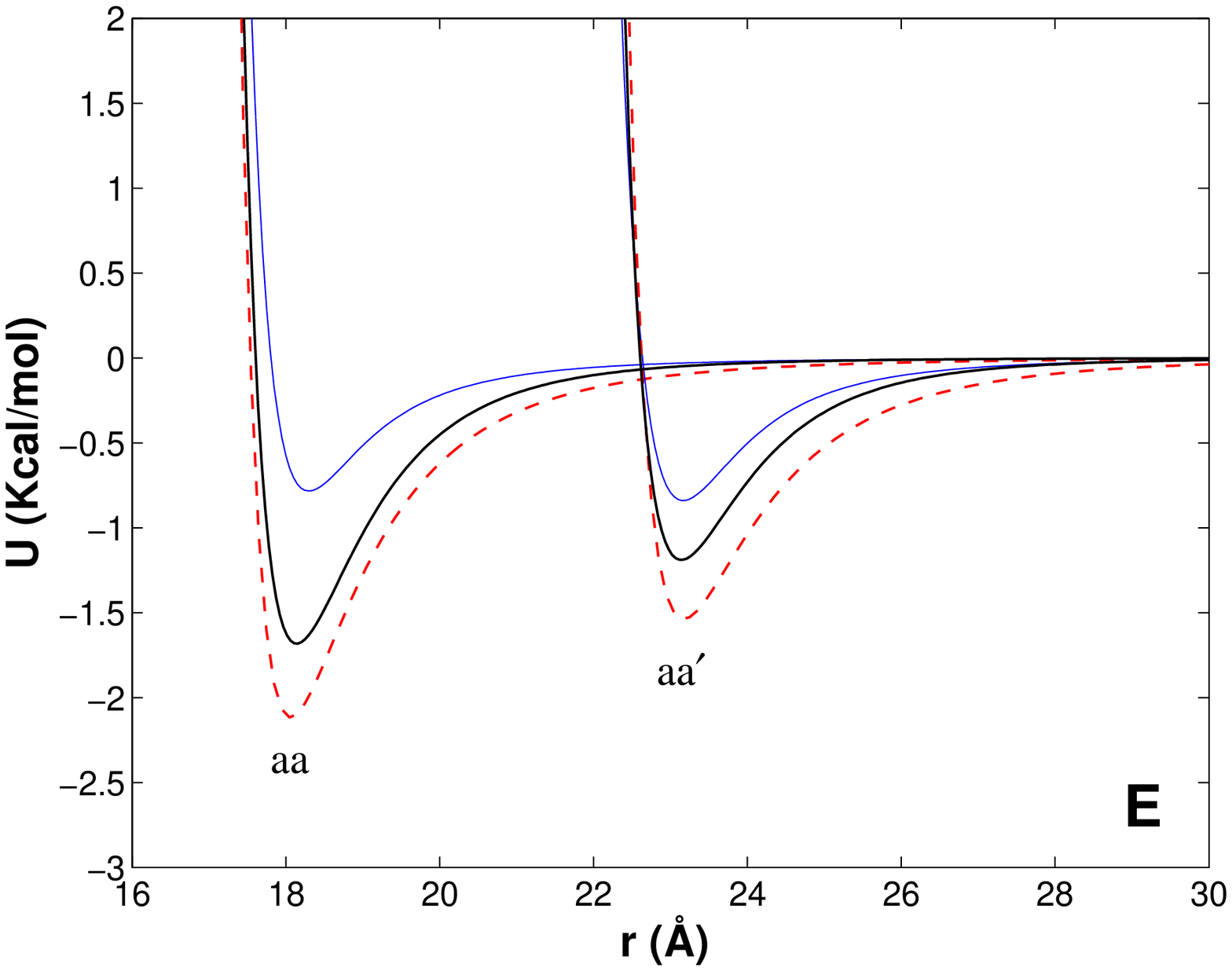}
\caption{The heterogeneous interaction between the pair perylene
(oblate) and sexithiophene (prolate) for the 18 orthogonal
configurations. The black thick lines denote the \resq potential,
the red dashed lines refer to the biaxial-GB~\cite{BFZ98} and the
reference atomistic summation (MM3~\cite{MM3}) is denoted by blue
thin lines. A combination of homogeneous interaction parameters
(Table~\ref{tab:param}) have been used without further optimization.
The error measures are: $\Omega_{RE^2}=6.5\times 10^{-3},
\Omega_{GB}=7.7\times 10^{-3}$. The graphs are grouped in five
plates as: side-by-side (A), cross (B), T-shaped 1 (C), T-shaped 2
(D) and  end-to-end (E) interactions and are labeled according to
the notation introduced in Table~(\ref{tab:ortho}).}
\label{fig:orthoplot}
\end{figure*}

The heterogeneous interaction between two molecules $\mathcal{M}_1$
and $\mathcal{M}_2$ is calculable by equations \eqref{eq:re2A} and \eqref{eq:re2R}
once the characteristic tensors of each molecule ($\mathbf{S}$ and $\mathbf{E}$)
along with the heterogeneous Hamaker constant $A_{\mathcal{M}_1 \mathcal{M}_2}$ and
the atomic potential radius $\sigma_{\mathcal{M}_1 \mathcal{M}_2}$ are available.
The heterogeneous Hamaker constant may be evaluated directly with a force-field
parameter lookup. Moreover, the arithmetic mean of $\sigma_{\mathcal{M}_1
\mathcal{M}_1}$ and $\sigma_{\mathcal{M}_2 \mathcal{M}_2}$ is a reasonable
estimate for the heterogeneous interaction radius, $\sigma_{\mathcal{M}_1\mathcal{M}_2}$.
Therefore, the homogeneous interaction parameters of the molecules $\mathcal{M}_1$
and $\mathcal{M}_2$ are sufficient to describe their respective heterogeneous interaction
using the \resq potential. Apparently, there is no trivial mixing rule available for
the energy scale of the biaxial-GB potential. Inspired by the atomic mixing rules
and the theory of the Gay-Berne potential, we have used Berthelot's geometric
averaging rule for this purpose. An instance of a heterogeneous interaction has been
illustrated in Fig.~(\ref{fig:orthoplot}) for the pair Perylene (oblate) and
Sexithiophene (prolate)~\cite{Mols}. The results are quite promising for a coarse-grained
model; However, further optimization will theoretically yield to superior results.
Concluding from the graphs, the \resq potential performs significantly better at
end-to-end and cross interactions compared to the biaxial-GB. The error measures
($\Omega_{RE^2}=6.5\times 10^{-3}, \Omega_{GB}=7.7\times 10^{-3}$) agree with this
observation.

\subsection{The advantages over practical atomistic implementations}
\begin{figure*}
\center
\includegraphics[bb=35 185 560 610, scale=0.4]{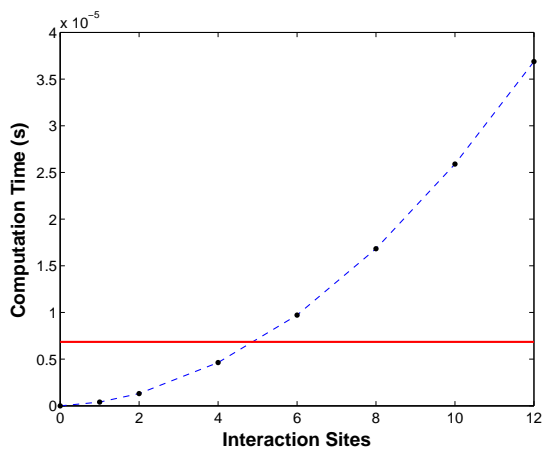}
\caption{The average computation time of exact LJ(6-12) atomistic
summation with respect to the average number of interacting sites (blue dashed line)
and \resq single-site potential (red continuous line).}
\label{fig:time}
\end{figure*}

As mentioned before, the atomistic evaluation of long-range
interaction potentials involve computationally expensive double
summations over the interaction sites, resulting in a quadratic time
cost with respect to the average number of interactions sites.
However, the average computation time of a single-site potential is
intrinsically constant, regardless of the number of interacting
atoms. These observations have been quantified in
Fig.~(\ref{fig:time}) which is a comparison between the computation
time of an exact LJ(6-12) atomistic summation and an efficient
implementation of the \resq potential~\cite{RE2CLib}. Concluding
from the graph, employing the \resq potential for molecules
consisting as low as~$\sim 5$ atoms (or $\sim 25$ overall atomic
interactions) is economic.

The atomistic summations are practically employed with a proper
large distance atomic cutoff in order to reduce the computation
time. In the presence of large distance cutoffs, long range
correction potential terms~\cite{Leach} are usually used to
compensate the submergence of particles beyond the cutoff distance.

Considerable errors may be introduced by choosing small atomic
cutoff distances compared to the dimensions of the interacting
molecules. Therefore, it is expectable that a CG model yield to
relatively better results compared to truncated atomistic summations
in certain configurations. A figurative situation is the end-to-end
interaction of two long prolate molecules. In such configurations,
usual atomic cutoffs ($\simeq 2.5\sigma$) can be small enough to
dismiss the interaction between the far ends of the molecules.
Moreover, long range correction terms are of little application in
this case due to the excessive inhomogeneity and the small number of
interacting particles.

\begin{figure*}
\center
\includegraphics[bb=35 185 560 610, scale=0.4]{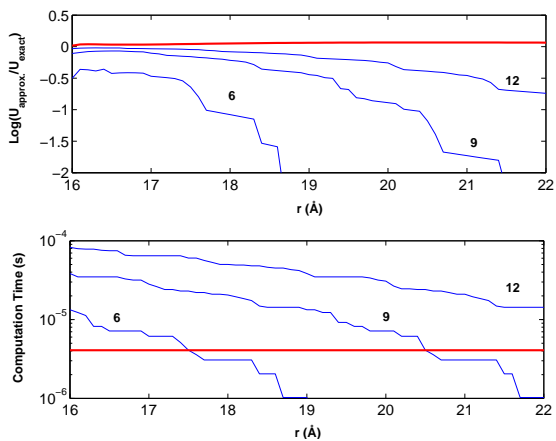}
\caption{A comparison between the truncated atomistic descriptions
with hard atomic cutoffs and the \resq potential for the end-to-end
interaction of the pair Pentacene. Thick lines denote the \resq
potential while thin lines represent atomistic summations with
different atomic cutoffs (6, 9 and 12 \AA). (A) Logarithmic relative
error of the \resq potential and truncated atomistic summations
(with respect to the exact atomistic summations) vs. center
separation. (B) Time consumption of different approximations vs.
center separation.} \label{fig:err_cut}
\end{figure*}

This effect has been illustrated in Fig.~(\ref{fig:err_cut}) for
Pentacene molecule~\cite{Mols}. The first panel is a semi-log plot
of the relative error for the \resq potential together with three
atomistic approximations with different cutoffs (6, 9 and 12 \AA).
The discontinuity of the truncated atomistic summations is a result
of hard cutoffs. In a proper atomic implementation, tapering
functions are used to avoid such discontinuities. Concluding from
the graphs, the CG description introduces less error in all ranges
of this configuration compared to the truncated atomistic
summations, even with unusually large atomic cutoff distance ($12
\AA$). Moreover, the evaluation of the CG interaction potential
requires a considerably lower computation time.

\section{Intermolecular Vibrations and Single-site Potentials}
In this section, we study the proposition of ideal rigidity of the
molecules, which is widely assumed in single-site approximations of
extended molecules, including our own study in the previous
sections. The samplings are usually taken from the relative
orientations of the unperturbed and geometrically optimized
structures. The resulting parameterization will be used in molecular
dynamics simulations in which internal vibrations may not be
negligible. We will introduce a method to estimate the error
introduced by this supposition in the first part of this study. A
parameterization based on the Potential of Mean Force (PMF) is
probably the best one can achieve with the coarse-grained models,
although the samplings are expensive. We will study such
parameterizations in the second part.

\subsection{Analysis of the Mean Relative Error}\label{subsec:err}
The PMF for the interaction of semi-rigid molecules in an arbitrary
ensemble may be expressed as an additive correction term to the the
interaction potential of the respective rigid molecules. We will
show that these correction terms are expressible in terms of
statistical geometric properties of a molecule in the ensemble. The
PMF between two molecules $\mathcal{M}_1$ and $\mathcal{M}_2$ with a
mean center separation of $\rl{\mathbf{r}}= \mathbf{r}_2 -
\mathbf{r}_1$ and mean orientation tensors $\mathbf{A}_1$ and
$\mathbf{A}_2$ is defined as:
\begin{equation}\label{eq:pmf}
U_{pmf}(\rl{\mathbf{r}}, \mathbf{A}_1, \mathbf{A}_2)=\left<U\left(\mathcal{M}_1,
\mathcal{M}_2\right)\right>
\end{equation}
where $\langle\cdot\rangle$ denotes the ensemble averaging. The mean
location of intermolecular particles are expected to remain
unchanged compared to the unperturbed structures for a large range
of temperatures as the internal structures of semi-rigid molecules
are mainly governed by harmonic bond stretching and angle bending
potentials. \\

We may assume the location of each particle as a random variable,
sharply peaked at its mean value. Therefore, we denote the location of
$i$th particle measured in its principal coordinate system by:
\begin{equation}
\mathbf{r}_i = \bar{\mathbf{r}}_i + \delta\mathbf{r}_i
\end{equation}
where $\bar{\mathbf{r}}_i = \left<\mathbf{r}_i\right>$ and
$\delta\mathbf{r}_i$ is a displacement due to internal vibrations
with vanishing average. The PMF of the interaction between the
molecules $\mathcal{M}_1$ and $\mathcal{M}_2$ is defined as:
\begin{equation}\label{eq:pmfLJ}
U_{pmf}(\rl{\mathbf{r}}, \mathbf{A}_1, \mathbf{A}_2)=
\left<\sum_{i\in\mathcal{M}_1}\sum_{j\in\mathcal{M}_2}U_a
\left(\|\mathbf{r}_i-\mathbf{r}_j\|;i,j\right)\right>
\end{equation}
where $U_a(\cdot)$ is the atomistic interaction potential. It is easy to
show that up to the second moments:
\begin{multline}\label{eq:aveR}
\left<\|\mathbf{r}_i-\mathbf{r}_j\|\right> \simeq
\|\mathbf{r}^0_{ij}\| + \\
\frac{1}{2\|\mathbf{r}^0_{ij}\|}\sum_{k=3}^3
\mathrm{Var}(\delta\mathbf{r}_i-\delta\mathbf{r}_j).\hat{\mathbf{e}}_k
\bigg(1-\frac{\big(\mathbf{r}^0_{ij}.\mathbf{e}_k\big)^2}{\|\mathbf{r}^0_{ij}\|^2}\bigg)
\end{multline}
where $\mathbf{r}^0_{ij} = \bar{\mathbf{r}}_i-\bar{\mathbf{r}}_j$.
We have neglected the covariance between the coordinates.
Using the last relation, we reach to a second-order estimate of Eq.~(\ref{eq:pmfLJ}):
\begin{multline}\label{eq:pmfLJexpanded}
U_{pmf}(\rl{\mathbf{r}}, \mathbf{A}_1, \mathbf{A}_2;
\mathcal{M}_1,\mathcal{M}_2) \simeq \\
\sum_{i\in\mathcal{M}_1}\sum_{j\in\mathcal{M}_2}U_a\left(\|\bar{\mathbf{r}}_i-
\bar{\mathbf{r}}_j\|;i,j\right)+\\
\frac{1}{2}\sum_{i\in\mathcal{M}_1}\sum_{j\in\mathcal{M}_2}\Bigg(\sum_{k=1}^3
\mathrm{Var}(\delta\mathbf{r}_i-\delta\mathbf{r}_j).\hat{\mathbf{e}}_k\times\\
\bigg(1-\frac{\big(\mathbf{r}^0_{ij}.\mathbf{e}_k\big)^2}{\|\mathbf{r}^0_{ij}\|^2}\bigg)
\frac{U_a'(\|\mathbf{r}^0_{ij}\|;i,j)}{\|\mathbf{r}^0_{ij}\|}+ \\
\sum_{k=1}^3\mathrm{Var}(\delta\mathbf{r}_i-\delta\mathbf{r}_j).\hat{\mathbf{e}}_k
\frac{\big(\mathbf{r}^0_{ij}.\mathbf{e}_k\big)^2}{\|\mathbf{r}^0_{ij}\|^2}
U_a''(\|\mathbf{r}^0_{ij}\|;i,j)\Bigg)
\end{multline}
The first term of the right hand side is the interaction energy of
the averaged structures, where the remaining terms are second-order
corrections. The RMS of the relative error introduced by neglecting
the correction terms (e.g. the error in parameterizations based on
unperturbed samplings) may be evaluated formally via the following integral:
\begin{equation}\label{eq:RMSint}
\mathcal{E}(T)=\left(\frac{1}{\mathcal{E}_0(T)}\int_{\omega\in\Omega(T)}
\exp{\left(-\frac{U(\omega)}{k_B T}\right)}\left[\frac{\delta
U(\omega)}{U(\omega)}\right]^2 dN_{\omega}\right)^{\frac{1}{2}}
\end{equation}
where $\omega$ is a relative orientation, $dN_{\omega}$ is
a differential measure of orientations near $\omega$, $\Omega(T)$
being the ensemble and $\delta U(\omega)$ is the
second-order correction defined by Eq.~(\ref{eq:pmfLJexpanded}).
$\mathcal{E}_0(T)$ is the normalization factor defined as:
\begin{equation}
\mathcal{E}_0(T) = \int_{\omega\in\Omega(T)}
\exp{\left(-\frac{U(\omega)}{k_B T}\right)}dN_{\omega}
\end{equation}

In practice, the spatial variance of each particle in an ensemble is
most easily obtainable through an MD simulation. Once the statistical
information are accessible, $\mathcal{E}(T)$ is most easily evaluated by
Monte Carlo integration. Neglecting the covariance between the dislocation
of the particles, we are implicitly overlooking the stretching and bending
of the molecules at close contact configurations. Although our proposed error
analysis disregards this phenomenon, it still measures the introduced error
due to purely thermal vibrations.\\

\begin{figure*}
\center
\includegraphics[bb=35 185 560 610, scale=0.4]{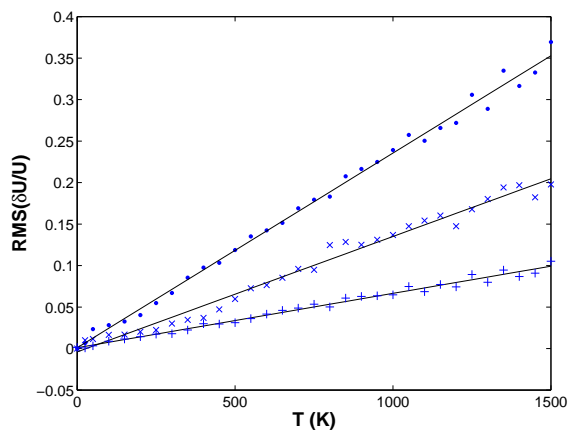}
\caption{Relative deviations from the PMF vs. temperature for three different
molecules. The signs indicate the MD simulation data. The continuous
lines are linear regressions. (1) Plus signs: Benzene (2) Cross
signs: Perylene (3) Dots: Sexithiophene.} \label{fig:err}
\end{figure*}

We have demonstrated this error analysis method for three different
molecules in a large range of temperatures. The statistical
information was extracted from several MD simulation snapshots with
the aid of TINKER molecular modeling package~\cite{TINKER}, each
with 32 molecules and with periodic boundary conditions in an NVT
ensemble (Fig.~\ref{fig:err}). For each isothermal ensemble, the RMS error
has been evaluated using the MC integration of Eq.~(\ref{eq:RMSint}) for $10^5$
random orientations. The relation between the RMS error and the temperature is
noticeably linear. The linear regression analysis has been given at Table~
(\ref{tab:err}). According to the required degree of precision, one
can define a trust region for the temperature using diagrams like
Fig.~(\ref{fig:err}). For example, a mean relative error of
less than 10\% is expected for temperatures less than 1500K in a homogeneous
ensemble of Benzene molecules, concluding from the graph. It is also concluded
that the studied prolate molecule (Sexithiophene) exhibits a higher relative error
due to its considerably higher elasticity, compared to the oblate molecules
(Perylene and Benzene).

\subsection{Parameterizations based on the Potential of Mean Force}
The error introduced by the assumption of ideal rigidity may not be
negligible for certain purposes, concluding from the previous
analysis. However, a coarse-grained potential which is parameterized
on a PMF basis is theoretically advantageous as it is expected to
describe the mean behaviors closer to the atomistic model.
Phenomenologically speaking, the internal degrees of freedom will
soften the repulsions at close contacts while the thermal vibrations
are expected to smoothen the orientation and separation
dependencies of the interaction.\\

The PMF for a given macroscopic orientation may be evaluated through
a Constrained Molecular Dynamics simulation (CMD) process with
appropriate restrains. We have used harmonic restraining potentials
for the center separation vector and on the deviations from the
desirable principal basis for each molecule in order to keep them at
the desired orientation. In order to reduce the random noise of the
evaluated PMF, we applied a fifth-order Savitsky-Golay smoothing
filter followed by a piecewise cubic Hermite interpolating
polynomial fitting to the PMF samples.\\

\begin{figure*}
\center
\includegraphics[bb=35 185 560 610, scale=0.4]{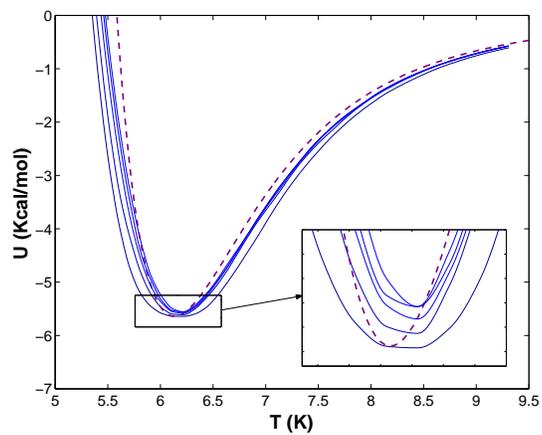}
\caption{Potential of Mean Force between the pair perylene for the
cross configuration {\bf bc}. The dashed line indicate the
interaction potential of the unperturbed structures while the
continuous lines, ordered descending with respect to their
well-depths, represent the PMF at temperatures 100K, 300K, 500K,
700K and 900K, respectively. } \label{fig:pmf}
\end{figure*}

Fig.~(\ref{fig:pmf}) is a plot of the evaluated PMF between the pair
perylene for the cross configuration {\bf bc}. The upper (and
interior) plots refer to higher temperatures. The expansion of the
potential well width at lower temperatures is related to the
tendency of the molecules to bend and stretch and thus,
resulting in a softer interaction while the shift of the soft contact and
potential well distance along with the elevation of the potential
well is associated to the thermal vibrations and hence, the
expansion of the effective volume of the molecules. The
temperature-dependant parameterizations (based on the evaluated PMF)
given at Table~(\ref{tab:param_pery}) justifies these qualitative
discussions. One may associate the contraction of the molecule
($\sigma_x$, $\sigma_y$ and $\sigma_z$) and the expansion of the
atomic interaction radius at lower temperatures to contraction of
the molecule at rough repulsions and widening of the potential well,
respectively. Furthermore, the expansion of the molecule volume at
higher temperatures reflect the overcoming of thermal vibrations to
the flexibility of the molecule. According to
tables~(\ref{tab:param}) and (\ref{tab:param_pery}), the overall
error measures ($\Omega$) for PMF parameterizations closely match
the same measure for the unperturbed structure. Thus, the same
functional form may be used to represent the PMF as well.

\section{Conclusion}
We have proposed and demonstrated a parameterization method for
the \resq anisotropic single-site interaction potential which
leads to a globally valid description of the attractive and repulsive
interaction between arbitrary molecules. Unlike the biaxial-GB~\cite{BFZ98},
the \resq potential gives the correct large distant interaction~(Fig.~\ref{fig:loglog})
while having a superior performance in the close contact region~(Fig.~\ref{fig:orthoplot}).
The Potential of Mean Force is representable with the same functional form
of the \resq potential. Compared to the parameterizations
given at~\cite{BFZ98}, The structure tensors agree significantly better
with the spatial distribution of the intermolecular particles.
It has also been shown that the coarse-gained models perform significantly
better at certain configurations in comparison to a truncated atomistic
summation with large distance cutoffs.

\section{Acknowledgment}
M. R. Ejtehadi would like to thank Institute for studies in Theoretical
Physics and Mathematics for partial supports.

\appendix

\section{The orientation dependent terms}
\label{sec:orientation}
We will briefly quote computable expressions for the orientation
dependant terms from the original article~\cite{EE03}. The term $\rl{\chi}$
quantifies the strength of interaction with respect to the local atomic
interaction strength of the molecules and is defined as:
\begin{equation}\label{eq:chidef}
\rl{\chi}(\mathbf{A}_1, \mathbf{A}_2, \rhat) =
2\rhat^T\rl{\mathbf{B}}^{-1}(\mathbf{A}_1, \mathbf{A}_2)\rhat
\end{equation}
where $\rl{\mathbf{B}}$ is defined in terms of the orientation tensors
$\mathbf{A}_i$ and relative well-depth tensors $\mathbf{E}_i$:
\begin{equation}\label{eq:Bdef}
\rl{\mathbf{B}}(\mathbf{A}_1, \mathbf{A}_1) = \mathbf{A}_1^T
\mathbf{E}_1 \mathbf{A}_1 + \mathbf{A}_2^T \mathbf{E}_2
\mathbf{A}_2.
\end{equation}
The term $\rl{\eta}$ describes the effect of contact curvatures of the
molecules in the strength of the interaction and is defined as:
\begin{equation}
\rl{\eta}(\mathbf{A}_1, \mathbf{A}_2, \rhat) =
\frac{\det[\mathbf{S}_1]/\sigma_1^2 +
\det[\mathbf{S}_2]/\sigma_2^2}{\big[\det[\rl{\mathbf{H}}]/(\sigma_1+\sigma_2)\big]^{1/2}},
\end{equation} \\
The projected radius of \textit{i}th ellipsoid along $\rhat$ ($\sigma_i$)
and the tensor $\rl{\mathbf{H}}$ are defined respectively as:
\begin{equation}\label{eq:sigdef}
\sigma_i(\mathbf{A}_i,
\rhat)=(\rhat^T\mathbf{A}_i^T\mathbf{S}_i^{-2}\mathbf{A}_i\rhat)^{-1/2}
\end{equation}
\begin{equation}\label{eq:Hdef}
\rl{\mathbf{H}}(\mathbf{A}_1,\mathbf{A}_2,\rhat) =
\frac{1}{\sigma_1} \mathbf{A}_1^T \mathbf{S}_1^2 \mathbf{A}_1 +
\frac{1}{\sigma_2} \mathbf{A}_2^T \mathbf{S}_2^2 \mathbf{A}_2 .
\end{equation}
There's no general solution to the least contact distance between
two arbitrary ellipsoids ($\rl{h}$). The Gay-Berne approximation~\cite{GB,
Perram96} is usually employed due to its low complexity and promising
performance:
\begin{equation}\label{eq:hgb}
\rl{h}^{GB} = \|\rl{\mathbf{r}}\| - \rl{\sigma}
\end{equation}
The anisotropic distance function $\rl{\sigma}$~\cite{BFZ98} is defined as:
\begin{equation}\label{eq:sig12def}
\rl{\sigma}=\left(\frac{1}{2}\rhat^T
\rl{\mathbf{G}}^{-1} \rhat\right)^{-\frac{1}{2}}
\end{equation}
where the symmetric overlap tensor $\rl{\mathbf{G}}$ is:
\begin{equation}\label{eq:Gdef}
\rl{\mathbf{G}} = \mathbf{A}_1^T\mathbf{S}_1^2\mathbf{A}_1 +
\mathbf{A}_2^T\mathbf{S}_2^2\mathbf{A}_2
\end{equation}

\bibliography{man}

\newpage

\begin{table*}
\center \caption{The 18 orthogonal configurations of two
dissimilar biaxial particles. A unitary operator ($\mathbf{U}$) followed by a
translation is applied to the second particle to reach the desired
configuration. We adopt the naming scheme introduced by Berardi {\it
et al}~\cite{BFZ98}. The operator $\mathbf{R}_e$ denotes a $\pi/2$ rotation with
respect to the axis $e$. A two-letter code is attached to each
configuration with respect to the faces perpendicular to
connecting vector of the ellipsoids. A prime is added if one or
three axes are antiparallel. Italic codes refer to configurations which
are degenerate in homogeneous interactions.}
\begin{tabular}{p{45pt} p{35pt} p{35pt} p{35pt}}
\\
$\mathbf{U}$ & $\rl{\mathbf{r}}\|\hat{\mathbf{e}}_{x1}$ &
$\rl{\mathbf{r}}\|\hat{\mathbf{e}}_{y1}$ &
$\rl{\mathbf{r}}\|\hat{\mathbf{e}}_{z1}$ \\
\hline \\
$\mathbf{I}$ & aa & bb & cc \\
$\mathbf{R}_z$ & ab & \textit{ba}$^\prime$ & cc$^\prime$ \\
$\mathbf{R}_y$ & ac$^\prime$ & bb$^\prime$ & \textit{ca}$^\prime$ \\
$\mathbf{R}_x\mathbf{R}_y$ & \textit{ac} & \textit{ba} & \textit{cb} \\
$\mathbf{R}_z\mathbf{R}_x\mathbf{R}_y$ & aa$^\prime$ & \textit{bc}$^\prime$ & cb$^\prime$ \\
$\mathbf{R}_x^T\mathbf{R}_z^T$ & ab & bc & ca \\
\end{tabular}
\label{tab:ortho}
\end{table*}

\newpage

\begin{table*}
\center \caption{\resq potential parameters for homogeneous
interactions of selected molecules~\cite{Mols}. The oblate molecules
are: (1) Perylene (2) Pyrene (3) Coronene (4) Benzene. The prolate
molecules are: (5) Sexithiophene (6) Pentacene (7) Anthracene (8)
Naphthalene (9) Toluene.}
\begin{ruledtabular}
\begin{tabular}{lccccccccc}
Mol. No. & $A_{12}$($10^2$ Kcal/mol) & $\sigma_c$(\AA) &
$\sigma_x$(\AA) & $\sigma_y$(\AA) & $\sigma_z$(\AA) & $E_x$ & $E_y$
& $E_z$ & $\Omega (10^{-3})$\\
\hline
Oblate: \\
$(1)$ & 36.36 & 3.90 & 4.20 & 3.12 & 0.49 & 3.96 & 2.39 & 0.49 & 9.6\\
$(2)$ & 28.36 & 3.91 & 4.24 & 3.07 & 0.45 & 3.98 & 2.35 & 0.43 & 9.9\\
$(3)$ & 21.01 & 3.83 & 4.27 & 4.26 & 0.54 & 2.84 & 2.85 & 0.35 & 12.5\\
$(4)$ & 84.95 & 3.99 & 2.14 & 1.82 & 0.36 & 4.60 & 3.70 & 1.03 & 6.8\\
\hline
Prolate: \\
$(5)$ & 49.44 & 4.07 & 10.96 & 1.99 & 0.46 & 6.30 & 1.16 & 0.35 & 10.7 \\
$(6)$ & 37.46 & 3.88 & 6.56 & 2.28 & 0.47 & 5.52 & 1.65 & 0.43 & 9.7\\
$(7)$ & 45.51 & 3.85 & 4.19 & 2.25 & 0.44 & 6.83 & 2.48 & 0.62 & 9.2\\
$(8)$ & 37.76 & 3.82 & 3.09 & 2.20 & 0.49 & 4.59 & 3.39 & 0.77 & 10.9\\
$(9)$ & 23.19 & 3.75 & 2.72 & 2.04 & 0.57 & 4.61 & 3.29 & 1.00 & 10.6\\
\end{tabular}
\end{ruledtabular}
\label{tab:param}
\end{table*}


\newpage

\begin{table*}
\center \caption{Linear regression analysis between $(\delta
U/U)_{RMS}$ and T for a few selected molecules~\cite{Mols}. The
molecules are: (1) Benzene (2) Perylene (3) Sexithiophene. The
linear relationship is defined as $(\delta U/U)_{RMS}$ = AT+B in all
cases.}
\begin{tabular}{p{50pt} p{70pt} p{60pt} p{60pt}}
\\
Mol. No. & A($10^{-4}K^{-1}$) & B($10^{-3}$) & $R^2$ \\
\hline
$(1)$ & 0.66 & 1.0 & 0.987\\
$(2)$ & 1.39 & -4.1 & 0.983\\
$(3)$ & 2.35 & 1.3 & 0.995
\end{tabular}
\label{tab:err}
\end{table*}

\newpage

\begin{table*}
\center \caption{\resq potential parameters for the homogeneous
interactions of the pair perylene at different temperatures.}
\begin{ruledtabular}
\begin{tabular}{ccccccccccc}
Temperature (K) & $A_{12}$($10^4$ Kcal/mol) & $\sigma_c$(\AA) &
$\sigma_x$(\AA) & $\sigma_y$(\AA) & $\sigma_z$(\AA) &
$E_x$ & $E_y$ & $E_z$ & $\Omega (10^{-3})$\\
\hline
100 & 14.59 & 4.83 & 3.77 & 2.69 & 0.10 & 3.56 & 2.13 & 0.41 & 15.2\\
300 & 2.04 & 4.51 & 3.97 & 2.85 & 0.24 & 3.52 & 2.18 & 0.43 & 14.1\\
500 & 1.21 & 4.40 & 4.04 & 2.91 & 0.30 & 3.49 & 2.20 & 0.44 & 12.6\\
700 & 0.99 & 4.33 & 4.09 & 2.97 & 0.32 & 3.53 & 2.19 & 0.44 & 12.4\\
900 & 0.82 & 4.31 & 4.09 & 2.98 & 0.35 & 3.43 & 2.15 & 0.44 & 12.6\\
\end{tabular}
\end{ruledtabular}
\label{tab:param_pery}
\end{table*}

\newpage

\newpage

\newpage

\newpage

\newpage

\newpage

\end{document}